\documentstyle[12pt,aaspp4]{article}
\lefthead{Abramenko et al.}
\righthead{Turbulent Dynamo}

\begin{document}

\title{Magnetic and Kinetic Power Spectra as a Tool to Probe the Turbulent
Dynamo}
\author{V. Abramenko, V. Yurchyshyn, and P.R. Goode
\affil{Big Bear Solar Observatory of New Jersey Institute of Technology, 40386, 
North Shore Ln, Big Bear City, CA, USA}
}

\begin{abstract}
Generation and diffusion of the magnetic field on the Sun is a key mechanism
responsible for solar activity on all spatial and temporal scales - from the
solar cycle down to the evolution of small-scale magnetic elements in the quiet
Sun. The solar dynamo operates as a non-linear dynamical process and is thought
to be manifest in two types: as a global dynamo responsible for the solar cycle
periodicity, and as a small-scale turbulent dynamo responsible for the formation
of magnetic carpet in the quiet Sun. Numerous MHD simulations of the solar
turbulence did not yet reach a consensus as to the existence of a turbulent
dynamo on the Sun. At the same time, high-resolution observations of the quiet
Sun from Hinode instruments suggest possibilities for the turbulent dynamo.
Analysis of characteristics of turbulence derived from observations would be
beneficial in tackling the problem. We analyze magnetic and velocity energy
spectra as derived from Hinode/SOT, SOHO/MDI, SDO/HMI and the New Solar
Telescope (NST) of Big Bear Solar Observatory (BBSO) to explore the
possibilities for the small-scale turbulent dynamo in the quiet Sun. 
\end{abstract}

\section{Introduction}
 
A dynamo process can be defined as the generation of the magnetic field from an
inverse cascade due to motions in a conducting medium. In the convective
envelope of the Sun, such motions are the result of turbulent convection in the
photosphere and beneath. The turbulent character of these motions is clearly
demonstrated by observations. As an example we present high-resolution
observations (Figure 1) of quiet sun (QS) granulation obtained recently with the
1.6 meter New Solar telescope (NST) of Big Bear Solar Observatory (BBSO,
Goode et al. 2010a,b). The corresponding two-hour movie is available at the BBSO
website {\tt http://bbso.njit.edu/gallery/nst\_tio\_20100803.mpg}.
Statistical analysis of this data set can be found in Abramenko et al. (2010, 2011) and at
{\tt http://www.bbso.njit.edu/{$\sim$}avi/2011NSF/}.

\begin{figure}[!ht]
\centerline{\epsfxsize=5.0truein \epsffile{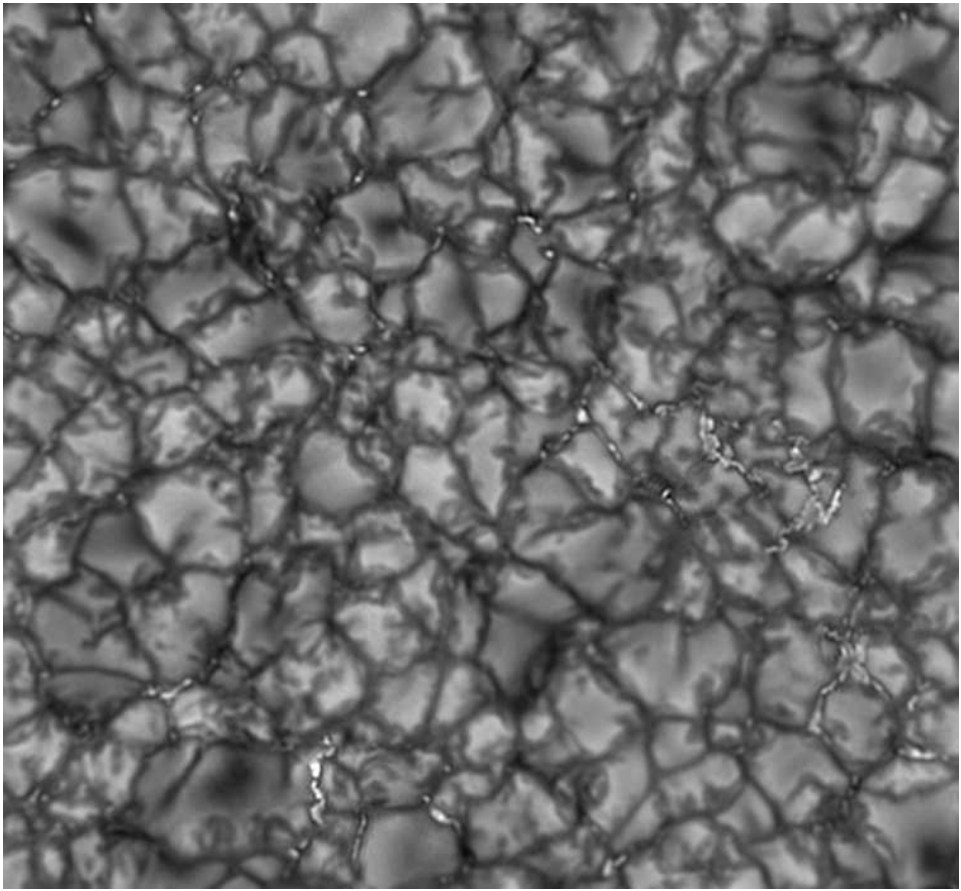}}
\caption{Image of solar granulation in a quiet sun area observed on August 3,
2010 at 17:06 UT with the NST of BBSO equipped with a TiO filter. The pixel size
is 0.0375 $''$ and the size of the image is 28.3 $\times$ 26.2 $''$. A movie of
the corresponding 2-hour data set further illustrates the turbulent nature of
solar plasma flows in the photosphere (see the text for the link).}
\end{figure}

Turbulent dynamo operation implies a gradual gain of magnetic energy inside a
volume (Batchelor 1950, Kazantsev 1968). Numerical simulations allow
one to probe the turbulent dynamo action (e.g., Boldyrev 2001,
Schekochihin et al. 2002, Cho et al. 2003, Boldyrev \& Cattaneo 2004). Indeed, fron an
initial moment for a simulation, one may track the evolution of the magnetic
energy accumulated in the volume. In the case an increase of magnetic energy is
detected, one can argue for a dynamo action. Unfortunately, this is not a
possibility when one deals with observational data. There is no apparent moment
of initiation, and the magnetic energy is constantly generated and dissipated
keeping the plasma in a state of dynamical equilibrium. We have to find some
approach to test this turbulent state and to make inferences on the
possibilities of turbulent dynamo action.

Two approaches might be suggested. First, one may test the interplay between the
kinetic and magnetic energies, i.e., to explore the spatial power spectra of
kinetic and magnetic energy. Second, one might test the structural organization
of the magnetic and velocity fields in QS, namely, to explore their
multifractal properties. Let us consider a theoretical background for the first
approach.

\begin{figure}[!ht]
\centerline{\epsfxsize=5.0truein \epsffile{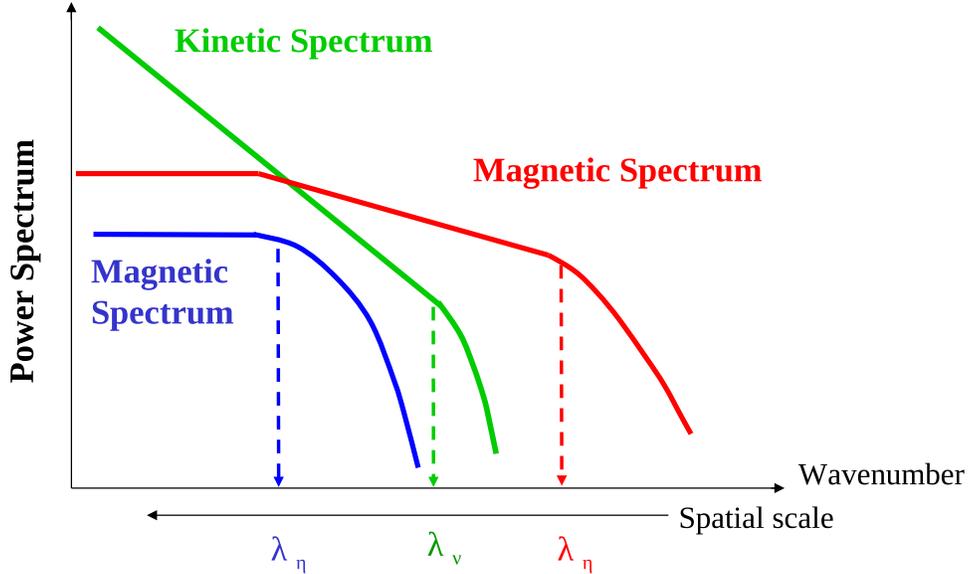}}
\caption{Sketch of a mutual behavior of the kinetic and magnetic power spectra
in a turbulent medium. Red and green lines: - a case of high Prandtl number
($Pr=\lambda_{\nu}/\lambda_{\eta}$) when the magnetic spectrum is more shallow
than the velocity spectrum, and at scales larger than $\lambda_{\nu}$, the
magnetic energy dominates the kinetic energy. In this limit, the situation is
favorable for turbulent dynamo action. Blue and green lines: - a case of low
Prandtl number, when the  magnetic spectrum shows a dissipation cut-off at
scales much larger than $\lambda_{\nu}$, magnetic eddies are exhausted at
smaller scales, and the dynamo action is questionable. }
\end{figure}

In Figure 2, the green line schematically represents a spatial power spectrum of
the turbulent velocities in the QS photosphere. The kinetic energy, being
deposited at large scales via the large-scale convective subphotospheric
motions, cascades down to smaller scales, producing the Kolmogorov-type spectrum
of -5/3. At some scale, $\lambda_{\nu}$, the kinetic vorticies start to
dissipate, which results as a cut-off of the spectrum. 

In order to generate magnetic fields at a scale $\lambda$, the following
condition should be met: the characteristic time for the curling of a magnetic
field line by a plasma's vortex (of size $\lambda$) should be shorter than the
characteristic time for the dissipation of a magnetic vortex of the same size.
This further requires that the  Prandtl number - the ratio of the kinetic
dissipation scale to the magnetic dissipation scale,
$\lambda_{\nu}/\lambda_{\eta}$, - has to exceed unity (Pikelner 1961,
Biskamp 1993). At a high Prandtl number the small-scale turbulent dynamo
action is both proved theoretically and in simulations (Batchelor 1950, 
Nakagawa \& Priest 1973, Boldyrev 2001, Schekochihin et al. 2002,
Cho et al. 2003). In this case, the magnetic spectrum (red line in Figure 2) is
shallower than the kinetic one (green line in Figure 2), so that magnetic
dissipation starts at scales, $\lambda_{\eta}$, much smaller than that for the
kinetic dissipation scale.

For the opposite case of low Prandtl number (blue line in Figure 2), the
magnetic dissipation starts on much larger scales, $\lambda_{\eta} >
\lambda_{\nu}$. Here, the turbulent dynamo action is very problematic, at least,
at very small scales ($\lambda < \lambda_{\eta}$) because these small magnetic
eddies cannot survive. However, there are studies aimed at probing the turbulent
dynamo at low Prandtl number via numerical simulations  (see, e.g.,
Boldyrev \& Cattaneo 2004, Iskakov et al. 2007, Pietarila Graham et al. 2010).

Thus, it is useful to see how the real kinetic and magnetic spectra
behave in the QS photosphere.

\section{Kinetic and Magnetic Energy Spectra}

\begin{figure}[!ht]
\centerline{\epsfxsize=6.5truein \epsffile{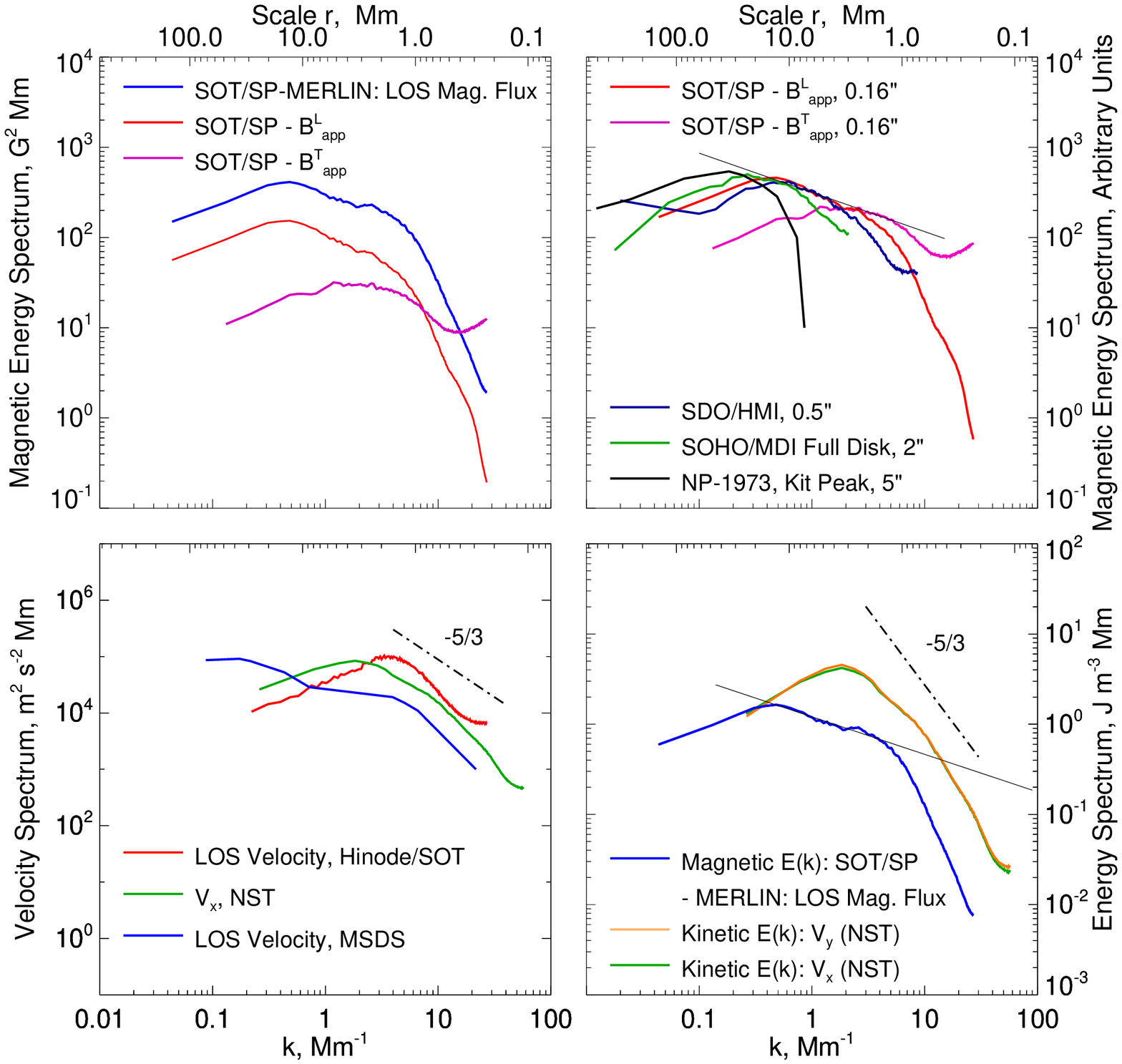}}
\caption{Magnetic and kinetic energy spectra from QS measurements. Top left
panel: magnetic spectra from Hinode/SOT/SP instrument 
(Tsuneta et al. 2008) recorded on 2007 March 10
at 11:37 UT. Top right: magnetic spectra derived from observations with
different instruments. The pixel size of each data set is noted. For better
comparison, the spectra are shifted along the vertical axis. Black line:
Nakagawa \$ Priest (1973) spectrum. As the resolution improves, the cut-off of
the spectra occurs at higher wave-numbers. The thin straight line shows a
tentative behavior of the spectra when the resolution is better than
0.16$''$. Bottom left: Velocity spectra. Blue line: Peak du Midi observations of
Doppler velocities with resolution of 0.8$''$ (Espagnet et al. 1993; this
spectrum was shifted along the vertical axis). Red line: Doppler velocities from
Hinode SOT/SP. Green line: a spectrum of the $V_x$ component of the transverse
velocities calculated from the NST data set presented in Figure 1. Bottom right:
kinetic energy spectra over-plotted with a magnetic energy spectrum. Thin black
line is a best linear fit to the magnetic spectrum above a cutoff. The
Kolmogorov-type spectrum (-5/3) is shown with the dash-dot lines. }
\end{figure}

To obtain the magnetic and kinetic energy spectra in the QS photosphere, we
utilized 2D images of the magnetic and velocity field components. 
From a squared 2D Fourier transform of an image, we calculated
the angle-integrated 1D power spectra. The routine is described in details in
Abramenko et al. (2001), Abramenko (2005a).  

For a comprehensive analysis, we explore how the magnetic and kinetic spectra
vary as the resolution of instruments improves. For that, we augmented our own
calculations of the spectra by those published in the past, when the available
spatial resolution was poor. Assuming a uniform spatial distribution of the mass
at any depth, we consider the power spectrum of the velocity field components to
be a representative of the kinetic energy spectrum.

Figure 3 shows magnetic and velocity spectra as derived from observations. In
the lower left panel, three velocity spectra for QS areas are shown:
\begin{itemize}
\item a Doppler velocity spectrum from Peak du Midi observations with resolution
of 0.8$''$ (Espagnet et al. (1993), blue line);
\item a Doppler velocity spectrum from
Hinode/SOT
observations in the spectral line 6301\AA\ recorded on March 10, 2008 (red
line), the pixel size of 0.16$''$;
\item a transverse velocity spectrum from NST observations of the solar
granulation
obtained from the data presented in Figure 1. The transverse velocity field
components, $V_x$ and  $V_y$, were computed via a local correlation tracking
(LCT) technique (Strous 1996, Abramenko et al. 2008). 
The FWHM of the Gaussian tracking window was $10\times10$ pixels (0.375$''$). To
obtain one velocity map, an
averaging over 12 time-steps (2 minutes) was undertaken. Total, 54 flow maps of
342$\times$342 nodal points each were computed. Angle-integrated power spectra
(Abramenko 2005a) from each $V_x, V_y$ map were calculated and
then averaged over 54 maps. Resulting spectra of $V_x$ and $V_y$ are shown in
the lower panels of Figure 3.
\end{itemize}

All the spectra show a smooth transition to the Kolmogorov-type regime with a
slope of -5/3 at scales of about 1.5-3 Mm. No well-pronounced tendency following
the improvement in the spatial resolution is observed.

Magnetic energy spectra obtained from observations are shown in the top panels
of Figure 3. The top left panel shows spectra obtained from Hinode/SOT/SP data
of a QS area observed on 2007 Mar 10. The spectrum derived from the inversion
technique (blue line, the LOS-flux spectrum) is very similar in shape to the
spectrum of $B^L_{app}$ (red line) derived from the Stokes profiles intensities
(Lites et al. 2008), however the LOS-flux spectrum possesses more magnetic power
at each scale. This means that the power distribution along the spectrum and,
therefore, the magnetic structuring, are captured equally by both techniques.
The difference in power arises from the LOS-flux values being systematically
higher. The spectrum of the transverse component, $B^T_{app}$ (purple line),
shows a presence of noise at small scales (a positive slope range at smallest
scales), which is not the case for the longitudinal component spectra. The
specific energy stored in the transverse component appears to be lower than that
in the longitudinal component. The most important point here is that the
transverse spectrum extends its shallow slope further toward small scales than
the longitudinal spectrum does. This means that the turbulent dynamo action is
better manifested in the transverse field generation.

Gradual improvements in our knowledge of magnetic power spectra occur with
the improvement of spatial resolution of solar instrumentation, and are
illustrated in
the top right panel of Figure 3, where we plot together many published and
computed spectra (all spectra are shifted along the vertical axis). Peak du Midi
data (Nakagawa \& Priest 1973) with 5$''$ resolution (black line) show a
smooth cut-off at scales about 20 Mm. MDI full disk data (green line) of 2$''$
resolution demonstrate a cut-off at about 10 Mm. The steeper slope below 10 Mm
might be an energy cascade signature, however, observations with higher spatial
resolution (dark blue line, HMI data) demonstrate an extension of the shallow
magnetic spectra toward smaller scales, down to 4-5 Mm. Observations with
Hinode/SOT further show that the magnetic spectrum extends with approximately
the same slope (of about -1/3) down to 1 Mm scale. This behavior is essentially
different from that we saw for the velocity spectra. We may speculate that
future observations with better resolution will allow us to obtain even more
extended shallow magnetic spectrum, in accordance with the thin straight line in
the top right panel of Figure 3. Compared to the velocity spectrum (bottom
right panel in Figure 3), the slope and the extend of the thin line may ensure a
possibility for prevalence of magnetic energy over the kinetic energy at small
scales, below approximately $\lambda_{TD} \approx 1 - 0.5$ Mm, and therefore, a
possibility for turbulent dynamo action. Note that the position of the
kinetic spectrum relative to the vertical axis strongly depends on the magnitude
of the photospheric plasma density. Here we accepted its value as $10^{-7}$ g
cm$^{-3}$. Further studies of photospheric plasma's conditions may bring some
corrections in the position of the kinetic spectrum, and, therefore, in
$\lambda_{TD}$.

\section{Intermittency and Multifractality Spectra of the Magnetic Field}

The multifractal organization of the magnetic field means that random strong
peaks in the vector field transport by a random flow
(say, the magnetic field vector in a turbulent electro-conductive flow) 
correspond to structural features such as magnetic flux ropes or thin sheets of
magnetic field lines. Thus, a weak seed magnetic field may grow exponentially, a
phenomenon known as a turbulent dynamo. The presence of multifractality is
physically compatible with turbulent dynamo action. The property of
multifractality is also frequently referred to as intermittency (see, e.g.,
Abramenko (2008) for a review of intermittency and multifractality in solar
phenomena). Numerical simulations (Cho et al. 2002) of the MHD
turbulence at high Prandtl number demonstrate highly intermittent magnetic field
organization. Simulations of turbulent magnetic reconnection (Matthaeus \& Lamkin 1986)
show that magnetically dominated modes (in the kinetic and magnetic spectra)
correspond to strong peaks in both magnetic and velocity fields. 

Observations from Hinode show that the photospheric magnetic field in QS is
intermittent: the blue line in the left panel of Figure 4 is a flatness
function (for definition, see, e.g., Frisch 1995, Abramenko 2005b,
Abramenko \& Yurchyshyn 2010) which we calculated from Hinode/SOT/SP
LOS magnetic flux map of a QS area observed on February 3, 2008. At small scales
(less than 4 Mm), the function increases as the scale decreases, which means a
multifractal organization and intermittency of the magnetic field. For
comparison, in the same figure we show a flatness function (green line) calculated 
for a typical active region (AR) from observations with lower
resolution. Intermittency in the magnetic field is also present, however, at
much larger scales (2-30 Mm). 

\begin{figure}[!ht]
\centerline{\epsfxsize=3.5truein \epsffile{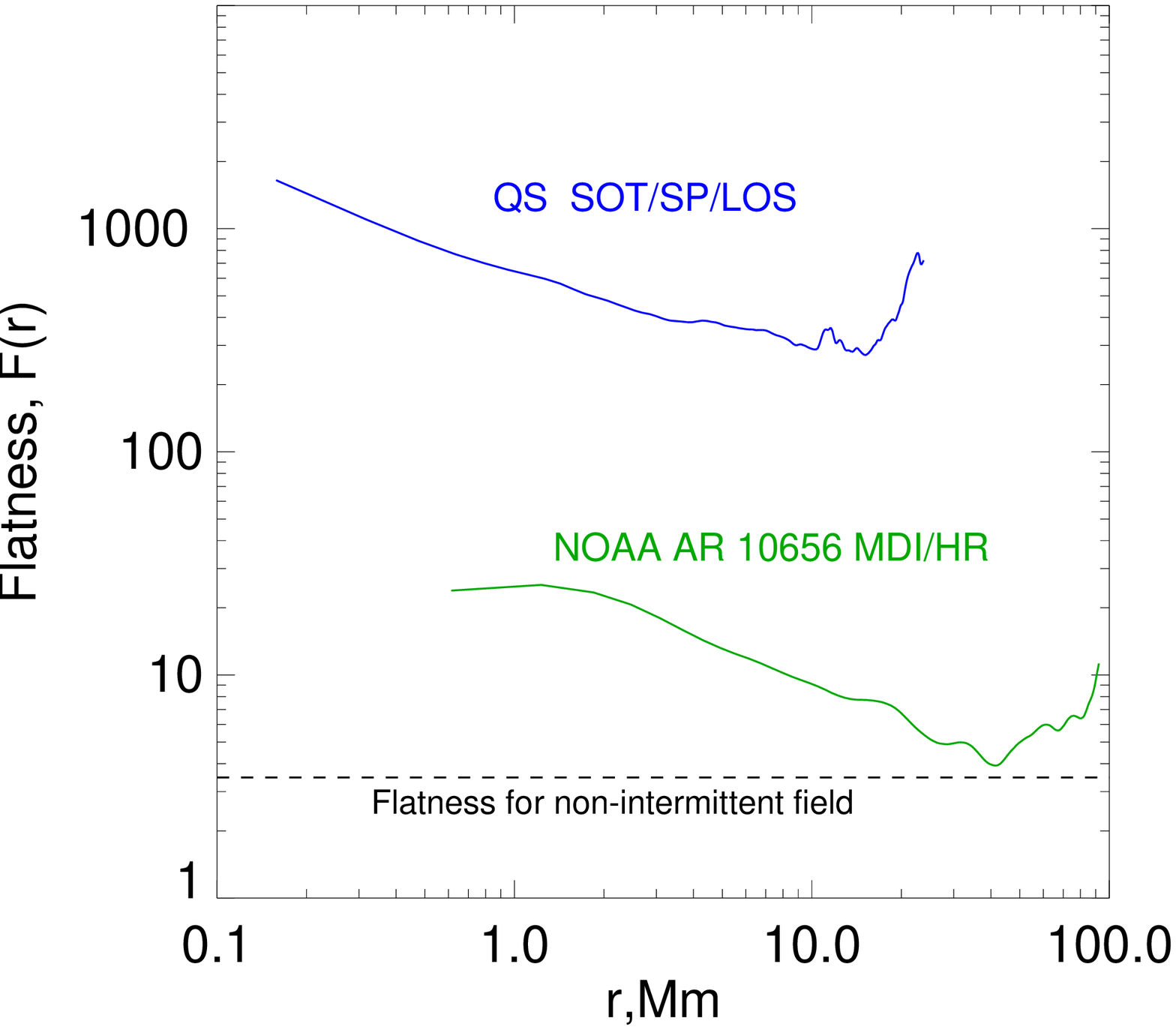}}
\centerline{\epsfxsize=3.5truein \epsffile{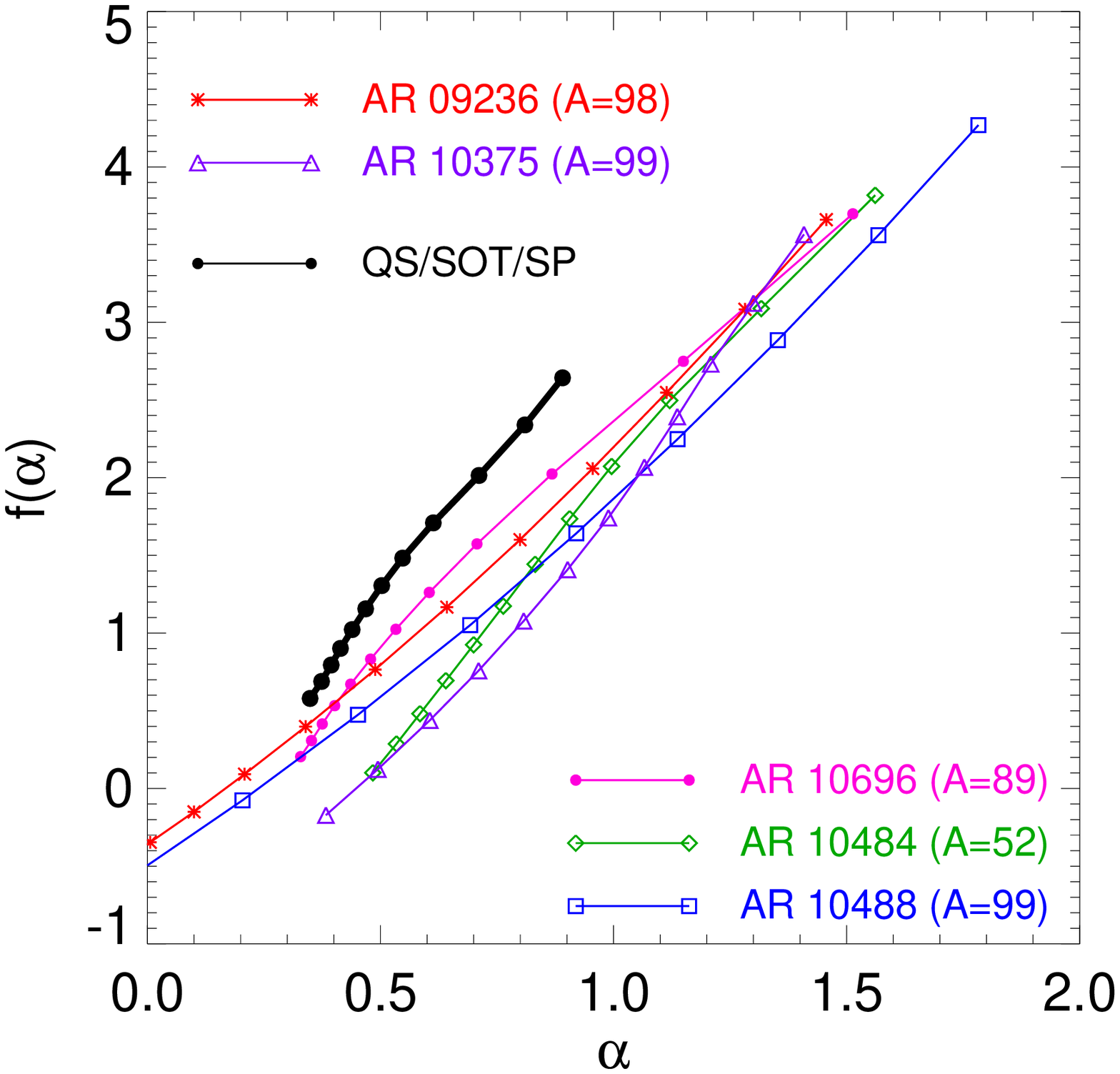}}
\caption{Top: Flatness functions of the LOS magnetic fields measured for a
typical QS area observed with Hinode/SOT/SP with a pixel size of 0.16$''$ and
for a typical AR (SOHO/MDI high resolution data, pixel size of 0.6$''$) Flatness
function is flat (dashed line) for a non-intermittent, monofractal, non-flaring
structure, and it becomes steeper with decreasing spatial scales for
intermittent fields. The steeper the function, the more complex the fields are.
Bottom: Multifractality spectra $f(\alpha)$ calculated for the same QS area and
for five ARs of high flaring activity (the flare index, $A$, proportional to the
flare productivity of an AR, is shown, Abramenko \& Yurchyshyn 2010).  }
\end{figure}

A widely accepted metric to characterize multifractality is the spectrum of
multifractality, $f(\alpha)$ (see, e.g., Feder 1989, Lawrence et al. 1993,
Schroeder 2000, Conlon et al. 2008, McAteer et al. 2010). 
Black line in the right panel of Figure 4 represents a spectrum of
multifractality, 
which we derived from the Hinode/SOT/SP LOS magnetic flux map of a QS area observed on
February 3, 2008. We see that a broad range of the scaling exponent $\alpha$ is
allowed, and for each $\alpha$, according to definition, there exists a sub-set,
i.e., a monofractal, of the fractal dimension $f(\alpha)$, so that all of them
being superposed, create a resulting observed magnetic multifractal. 
Behavior of the QS multifractality spectrum is similar to that observed for
strong-flaring ARs (see the right panel of Figure 4). 
This implies similar multifractal organization of the magnetic field, but in the
case of the QS it occurs
at much smaller spatial scales. 

\section{Summary}

New generation solar instruments, namely, Hinode/SOT/SP and BBSO/NST 
demonstrate previously unclear tendencies in the behavior of QS magnetic and
velocity fields. 

First, as the resolution improves, the shallow magnetic energy spectrum  tends
to extend toward higher wave-numbers (smaller scales), the slope remaining the
same, whereas the velocity spectrum shows a steady maximum at 1-3 Mm followed by
the Kolmogorov-type regime of $E(k) \sim k^{-5/3}$. This tendency necessarily
results in setting in high-Prandtl number turbulence with necessary turbulent
dynamo action at small scales below $\lambda_{TD} \approx 1 - 0.5$ Mm.  

Second, improved resolution of solar data revealed a highly
intermittent and multifractal nature of the small-scale magnetic field in QS
areas. This property of the magnetic field is extremely favorable for the
turbulent dynamo action at all scales where the intermittency/multifractality
is present.

The above results were presented and discussed at the 2010 Hinode-4 Meeting in Palermo 
(Italy). VA is thankfull to Yukio Katsukawa, Jan Stenflo, Saku Tsuneta and Bruce Lites
for helpful discussions. 

We gratefully acknowledge help of the NST team and support of NSF (ATM-0716512 and
ATM-0847126), NASA (NNX08AJ20G, NNX08AQ89G, NNX08BA22G), AFOSR
(FA9550-09-1-0655).
Hinode is a Japanese mission developed and launched by ISAS/JAXA,
collaborating with NAOJ as a domestic partner, NASA and STFC (UK) as
international partners. It is operated by these agencies in co-operation with
ESA and NSA (Norway).

{}

\end{document}